# End-of-Use Core Triage in Extreme Scenarios Based on a Threshold Approach

**Abstract ID: 309**
**Saeed Z.Gavidel**
**Wayne State University**
**Detroit, MI**

**Dr. Jeremy L. Rickli**
**Wayne State University**
**Detroit, MI**

## Abstract

Remanufacturing is a significant factor in securing sustainability through a circular economy. Sorting plays a significant role in remanufacturing pre-processing inspections. Its significance can increase when remanufacturing facilities encounter extreme situations, such as abnormally huge core arrivals. Our main objective in this work is switching from less efficient to a more efficient model and to characterize extreme behavior of core arrival in remanufacturing and applying the developed model to triage cores. Central tendency core flow models are not sufficient to handle extreme situations, however, complementary Extreme Value (EV) approaches have shown to improve model efficiency. Extreme core flows to remanufacturing facilities are rare but still likely and can adversely affect remanufacturing business operations. In this investigation, extreme end-of-use core flow is modelled by a threshold approach using the Generalized Pareto Distribution (GPD). It is shown that GPD has better performance than its maxima-block GEV counterpart from practical and data efficiency perspectives. The model is validated by a synthesized big dataset, tested by sophisticated statistical Anderson Darling (AD) test, and is applied to a case of extreme flow to a valve shop in order to predict probability of over-capacity arrivals that is critical in remanufacturing business management. Finally, the GPD model combined with triage strategies is used to initiate investigations into the efficacy of different triage methods in remanufacturing operations.

## Keywords
Remanufacturing; Sustainability; Triage; Generalized Pareto Distribution (GPD)

## 1. Introduction
Cores, as the material source of remanufacturing operations, play a significant role in the sustainability of this business, such that "*Who owns the core, owns the market*" [1]. In addition to the importance of core acquisition in remanufacturing, core pre-processing and sorting tasks play a pivotal role, such that they are rated as a critical remanufacturing activity by remanufacturers [2]. There are many investigations regarding core acquisition and sorting management in literature, however, the majority of investigations are focused on normal situations where there are no extreme patterns such as huge core arrivals to a remanufacturing facility. Galbreth and Blackburn [3] used normally distributed demand patterns for modelling optimum acquisition quantities, Aras et al. [4] modelled both customer demand and product returns as Poisson Processes for hybrid manufacturing and remanufacturing systems, and Teunter and Flapper [5] assumed normal distribution for demands to model optimum core acquisitions. Extreme core flow in remanufacturing/ recalls have been reported and so deserves in-depth investigations [6, 7].

Less likely extreme events that have extreme effects can be analyzed by Extreme Value Theory (EVT) [8]. Gavidel and Rickli [9] modelled extreme core flow to remanufacturing facilities. They used the Block Maxima (BM) approach of EVT, to model the extreme core flow to a valve shop of an industrial complex. Their results showed that application of EV model as a complementary tool to central tendency model suggests more efficient predictions [9]. One main objective in this work is improving model efficacy by switching from less efficient BM model to more efficient model developed based on threshold approach of EV theory. Although BM provides a reliable framework to model and analyze EV scenarios, it does not necessarily offer optimal solutions for all situations especially where the cost of business monitoring, data harvesting, and storage is high. Modelling based on BM and deriving associated Generalized Extreme Value (GEV) distributions (type I, II and III) needs large amount of data. An alternative for dealing with EV scenarios is threshold approach in which instead of monitoring the entire process to screen maxima of blocks, only Peaks Over Threshold (POT's), are recorded and analyzed [9,10]. In this investigation, we model the extreme core flow to remanufacturing facilities by threshold approach and show that for practical applications, such as remanufacturing, this approach yields better technical, analytical, and predictive results than BM approach. Our motivations to conduct the current investigation are both technical and practical. Developing a model based on BM requires large samples else estimation uncertainties can be large [11]. Since BM approach uses maxima of blocks to develop the GEV model and since there is just one maxima per block, other non-maximal but extreme values are disregarded.

Proper threshold determination is a vital step of the threshold approach and requires domain expert engagement, inspection and verification of data to understand, assess, and endorse the fitted model. Hence, threshold determination is expert oriented task, such that high levels of process knowledge and experience play important role [12]. The significant role of practitioners and engineering evaluations means that GPD is not a pure statistical approach. It should be noted that some methods have been developed to assess the threshold value through analytical methods [13]. In this research, we have used a common statistical approach to estimate threshold. Scarcity of process data can be challenging in some remanufacturing facilities, this scarcity can be a result of reasons such as; new facilities where there is no sufficient amount of long-run process data, process monitoring is hard or expensive. Hence, these facilities are not capable of providing sufficient amount of effective data to BM and may only be able to provide small samples. Table 1 presents a summary of motivations, behind GPD as our new modelling approach.

Table 1: Block maxima approach vs. threshold approach in EV scenarios

| Motivation | **Block Maxima-BM** | **Threshold** |
|---|---|---|
| **Technical** | Statistical and less practical, easy to apply [11] | Practice-oriented, less easy to apply [11] |
| | All the process data must be recorded | Just POT's must be recorded |
| | Non-critical maxima may be considered as extreme | Over thresholds are considered as extreme [15] |
| | No cutoff between normal and extreme pattern | Threshold can be used as a cutoff between normal and extreme modes |
| | Needs large samples, small samples lead to high prediction variations | Models small samples efficiently |
| **Analytical** | Less accurate model fitting behavior | Highly accurate model fitting behavior |
| | Wasteful if other approaches applicable [8] | Uses information efficiently [16] |
| | No threshold | Threshold determination is very challenging |

The remainder of this paper is organized as follows: section 2 presents statistical and analytical foundations of EVT for threshold approach. A solid synthesized dataset is used to develop the model and the developed GPD model is compared with its BM counterpart. In section 3, a dataset of extreme valve arrival to a valve shop of a chemical complex is investigated by threshold approach. The GPD model is built and parameters estimated and verified by proper statistical techniques. Parametric bootstrapping is used to evaluate model variability and statistical accuracy.

## 2. Methodology

Threshold approach for EV scenarios is based on a threshold value, $u$, and the exceedances (surpasses) over threshold (also known as POT's). Since the method is based on POT's, these values are of great concern and must be monitored and recorded. The theory of GPD, introduced by Balkema, de Haan and Pickands [10], maintains that for a vast class of underlying distributions, say $F$, the conditional distribution function of POT's, $F_u(x)$, for sufficiently large threshold, is well approximated by generalized Pareto distribution [8, 10].

$$G_{\xi,\sigma}(x) = \begin{cases} 1 - (1 + \xi \frac{x}{\sigma})^{-\frac{1}{\xi}}; & \xi \neq 0 \\ 1 - \exp\left(-\frac{x}{\sigma}\right); & \xi = 0 \end{cases} \quad (1)$$

Where $\xi$ and $\sigma$ are shape and scale parameters respectively. Having the unknown, cumulative distribution function $F_X(x)$, of a random variable, $x$, which describes the random core extreme arrival, the distribution of POT's for threshold like $u$, can be represented as a conditional distribution:

$$F_u(y) = P(X - u \leq y | X > u) = \frac{F_X(x) - F_X(u)}{1 - F_X(u)}, y = x - u > 0 \quad (2)$$

Applying approximation, we will have the following relation, $\lim_{x \to x_F} sup |F_u(y) - GPD_{\xi,\beta}(y)| = 0$, where $x_F$, is the right endpoint of the distribution. The excess GPD is as follows:

$$GPD_{\xi,\beta}(y) = \begin{cases} 1 - \left(1 + \xi \frac{y}{\beta}\right)^{-\frac{1}{\xi}}; \xi \neq 0 \\ 1 - \exp\left(-\frac{y}{\beta}\right); \xi = 0 \end{cases} \quad (3)$$

Where $y = x - u$ is a random variable as peak over threshold (POT), $\xi$ and $\beta$ are shape and scale parameters respectively. The threshold can be evaluated based on the mean of the GPD: if $Y$ is a random variable following a GPD with parameters $\sigma$ and $\xi$, then the expected value for $Y$, $E(Y)$, can be evaluated by (5) when $\xi < 1$, [8].

$$E(Y) = \frac{\sigma}{(1 - \xi)} \quad (4)$$

The issue of threshold determination is like determination of block size in maxima block approach and is a balance between bias and variance. It should be noted that any threshold greater than determined threshold can be used [13]. In this investigation we have used Eq. 4, to assess the threshold value. Prior to applying to real dataset, the GPD model is applied to a synthesized hurricane dataset, to observe its behavior under standard simulated datasets. It should be noted that the hurricane dataset is a part of large data repository provided by Information Technology Laboratory, National Institute of Standards and Technology (NIST) generated to simulate extreme value situations [17]. Fig. 1 shows Cumulative Distribution Functions (CDF) for corresponding BM, Normal and GPD for the hurricane dataset. It can be observed that GPD fits the ECDF better than BM and Normal distributions. Ding and et.al, have reported this superior fitting performance of GPD by applying statistical tests such as Kolmogoroff-Smirnov (K-S), correlation coefficient (R) and mean square error (MSE) in extreme behavior of Chinese rivers [18]. It is observable that both models derived by Maximum Likelihood (ML) and Probability Weighted Moments (PWM) parameter estimation techniques are consistent. However, this is only for large samples, for small samples PWM has better estimation performance [19]. Note that the evaluated threshold is 80.

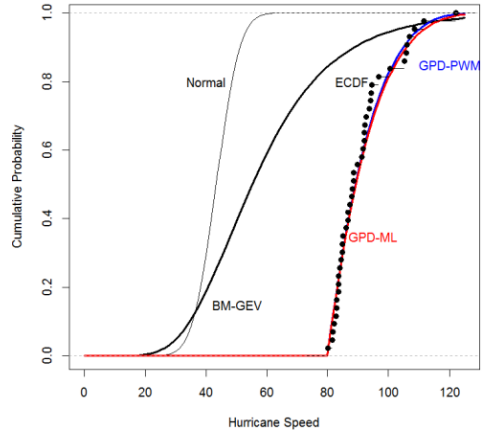

Figure 1: GPD demonstrates efficient fitting behavior

## 3. Case Study: Valve Remanufacturing

### 3.1. Parameter Estimation and Model Development

As a case study, consider arrivals of valves as cores in valve industry repair (a synonym for remanufacturing in this instance [20]) to the valve shop of a chemical complex for a period of 476 consecutive operational days [9]. The estimated mean remanufacturing capacity of valve shop is 10 units of valves per operational day. It should be noted that the threshold is evaluated by the methodology presented in this paper and considering the capacity of the valve shop, estimated threshold is far beyond the capacity of the valve shop. The 476 day consecutive time interval was considered as 158 three-day sub-intervals. Table 2 presents the estimated shape and scale parameters using PWM, due to small sample size. The Anderson-Darling (AD) statistical goodness of fit test is applied and *0.3022* is calculated that indicates a good fit. High *p* and AD-values strongly confirm that at common significance levels, such as like 0.1 and 0.05, efficient performance of GPD model cannot be rejected [21].

Table 2: Estimated GPD parameters for valves arrival

| Statistics | **Threshold** | Shape($\xi$) | Scale ($\beta$) | AD | p-value |
|---|---|---|---|---|---|
| Value | 49 | 0.1215 | 22.48 | 0.3022 | 0.9354 |

Scatter plot, GPD, BM-GEV and ECDF for flow of valves to the valve shop are presented in Fig.2. Similar to synthesized hurricane dataset, it can be seen that developed GPD fits the ECDF better than BM-GEV and this fact is observable as a fitting gap between BM-GEV and ECDF.

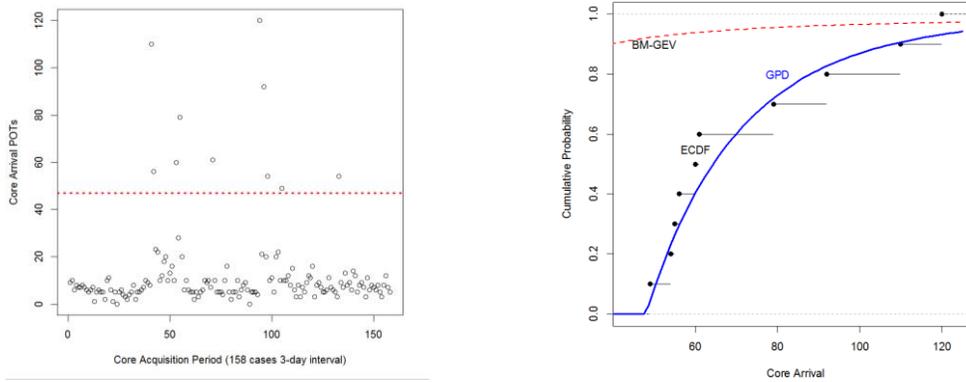

Figure 2: (a) Extreme core arrival scatter plot (b): Fitted GPD, BM-GEV and ECDF

### 3.2. Model Variation and Statistical Analysis

In this investigation, we have used *bootstrapping* as a tool to analyze statistical accuracy of our model. According to Bishop [22], the *statistical accuracy* of estimates can be evaluated by the variability of predictions between different bootstrapped datasets. Due to good fitness, indicated by the AD statistics, the model is efficient to conduct parametric bootstrapping [23]. In parametric bootstrapping, parameters of assumed population distribution are estimated and re-sampling is done by pseudo-population defined by the assumed distribution [14]. In this investigation, 2100 bootstrapped (parametric) datasets are generated from original GPD. This number of datasets are sufficiently higher than what proposed as minimum [23]. Two curves of total 2100 curves, have maximum deviation from the original model and Fig. 3, demonstrates ECDF, original GPD, and these two bootstrapped curves. As expected, the fitted GPD for original dataset is moderate and between two boundary bootstrapped curves.

The prediction capability for different levels of extreme core arrival by ECDF, original GPD, and marginal bootstrapped GPD's is visualized by a radar plot and presented in Fig.4, for both synthesized and real datasets. The values on the vertices are valve arrival and hurricane speeds. Probability of occurrence is represented by polygons as counterplots. It can be observed that the general trends of two plots is similar. It should be mentioned that tighter variations for hurricane dataset is due to better parameter estimations that is in turn is the result of greater sample size. As this case study demonstrates, both BM and GPD approaches defeat traditional central tendency approach, such as Normal distribution in extreme core arrival scenarios. However, GPD has shown technical and analytical advantages with respect to BM, this fact confirmed by synthesized dataset and observed in real dataset also. It has been shown that modeling by GPD is of special interest in situations where efficient utilization of information resources is of great concern. Finally, both GPD and BM are proposed as complimentary to central tendency approaches to analyze extreme situations.

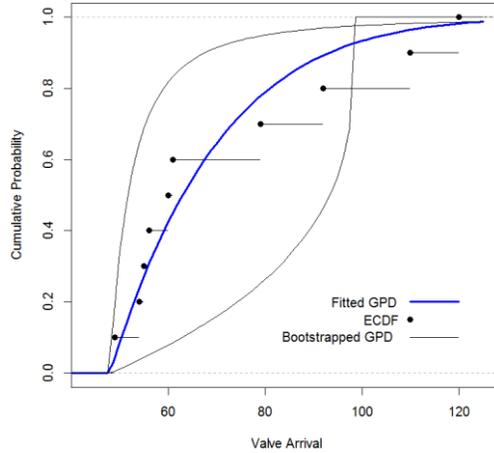

Fig 3: ECDF, Original GPD, Bootstrapped GPD

In general, conservative bootstrapped GPD, results in overestimation while the non-conservative bootstrapped GPD underestimates the occurrence probability of extreme events for some low values and then overestimates for some high values. It should be noted that maximum and minimum absolute errors belong to non-conservative bootstrapped model.

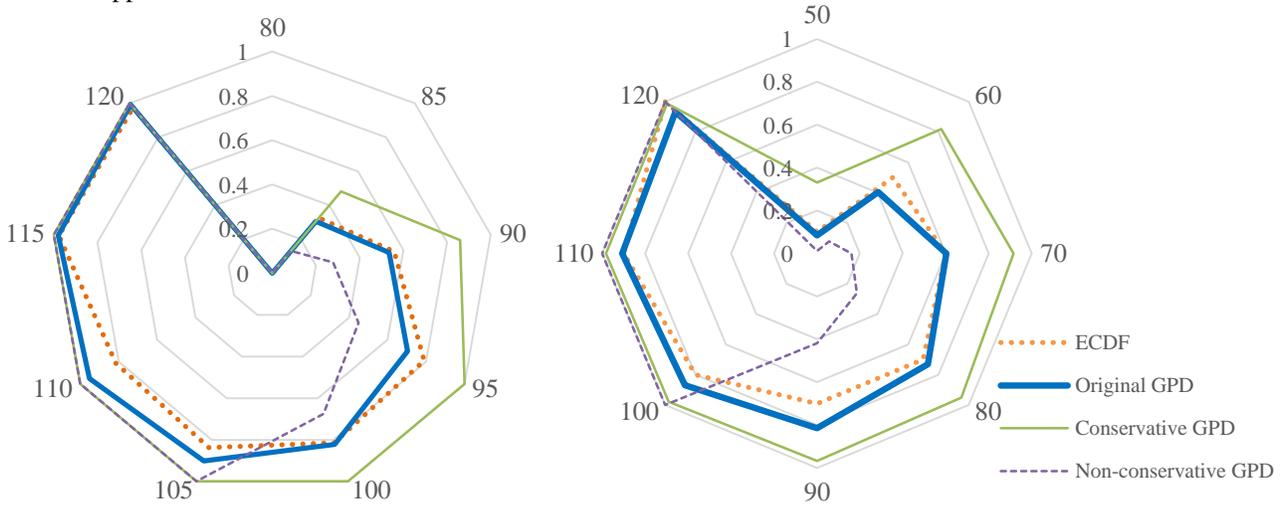

Figure 4: Statistical accuracy analysis of GPD model for Hurricane dataset (left) and Valve Arrival dataset (right)

## 4. Conclusion and Future Research

Central tendency approach is not the sole method to deal with core stream to remanufacturing facilities. In extreme situations, EV distributions both BM-GEV and GPD outperform central tendency counterparts such as normal distribution. Hence, EV models both BM-GEV and GPD can be used as a complementary tools along with central tendency approaches to analyze core stream to remanufacturing facility more efficiently. However, since in most of the industrial applications, we are just interested in extremes over pre-defined technical thresholds so entire process monitoring may be considered as wasteful. Threshold approach just uses POT's so if modeling with GPD is possible using other approaches is considered wasteful. On one hand GPD model fitted to experimental data showed appropriate fitting behavior such that AD goodness of fit test yield value of *0.3022* (equivalently p-value 0.9354) that means the hypothesis of Pareto model for valve arrival dataset cannot be rejected at common confidence levels. On the other hand, statistical evidences demonstrated that developed GPD model defeated both central tendency and BM-GEV models from analytical perspective also. As a contribution to remanufacturing from practical scope, developed model can be used to more efficiently model and analyze the extreme events in remanufacturing facilities especially, where facilities suffer from process data shortages.

EV models for core stream accompanied by proper model for quality, presents opportunities to triage cores in remanufacturing facilities. The model for the quality may be a wide spectrum of statistical scenarios such as uniform, binomial, normal or even EV distributions. EV model for quantity accompanied by a quality model can be used to develop a joint quantity-quality joint model to use in triaging remanufacturing cores based on calculated payoffs resulted from various triage strategy. To solve core triage based on proposed strategies, analytical techniques, computational and simulation techniques, grant opportunities for more future research.